\documentclass[oneside,12pt]{article}

\usepackage{geometry}
\usepackage{soul}
\usepackage{amsmath}
\usepackage{amssymb,amsthm,nicefrac}
\usepackage{textcomp}
\usepackage{amsfonts}
\usepackage{authblk}
\usepackage{algorithmic}
\usepackage{multirow}
\usepackage{colortbl}
\usepackage{color}
\usepackage[table]{xcolor}
\usepackage{epigraph}
\usepackage{caption}
\usepackage{subcaption}
\usepackage{hyperref}
\usepackage{tabularx}
\usepackage{float}
\usepackage{longtable}
\usepackage[pdftex]{graphicx}
\usepackage{setspace}
\usepackage{breakcites}
\usepackage{calc}
\newsavebox\CBox
\newcommand\hcancel[2][0.5pt]{%
  \ifmmode\sbox\CBox{$#2$}\else\sbox\CBox{#2}\fi%
  \makebox[0pt][l]{\usebox\CBox}%
  \rule[0.5\ht\CBox-#1/2]{\wd\CBox}{#1}}

\DeclareMathOperator*{\argmin}{argmin}

\newtheorem{Lemma}{Lemma}

\newtheorem*{spf*}{Single-player formulation}

\newtheorem{theorem}{Theorem}[section]

\theoremstyle{definition}
\newtheorem{definition}[theorem]{Definition}
\theoremstyle{definition}

\usepackage{algorithmic}
\usepackage{algorithm}

\title{Identifying Dynamic Regulation with Adversarial Surrogates}

\date{\vspace{-5ex}}
\author[1,2]{Ron Teichner} 
\author[3,2]{Naama Brenner}
\author[1,2]{Ron Meir}
\affil[1]{Viterbi Dept. of Electrical \& Computer Eng., Technion, Israel Institute of Technology}
\affil[2]{Network Biology Research Lab, Technion, Israel Institute of Technology}
\affil[3]{Dept. of Chemical Eng., Technion, Israel Institute of Technology}

\begin{document}
\maketitle
%

%





\begin{abstract}
  Homeostasis, the ability to maintain a stable internal environment in the face of perturbations, is essential for the functioning of living systems. Given observations of a system, or even a detailed model of one, it is both valuable and extremely challenging to extract the control objectives of the homeostatic mechanisms. Lacking a clear separation between plant and controller, frameworks such as inverse optimal control and inverse reinforcement learning are unable to identify the homeostatic mechanisms. A recently developed data-driven algorithm, Identifying Regulation with Adversarial Surrogates (IRAS), detects highly regulated or conserved quantities as the solution of a min-max optimization scheme that automates classical surrogate data methods. Yet, the definition of homeostasis as regulation within narrow limits is too strict for biological systems which show sustained oscillations such as circadian rhythms. In this work, we introduce Identifying Dynamic Regulation with Adversarial Surrogates (IDRAS), a generalization of the IRAS algorithm, capable of identifying control objectives that are regulated with respect to a dynamical reference value. We test the algorithm on simulation data from realistic biological models and benchmark physical systems, demonstrating excellent empirical results.
\end{abstract}

\section{Introduction}

Living systems maintain stability against internal and external perturbations, a phenomenon known as homeostasis \cite{billman2020homeostasis,hsiao2018control,kotas2015homeostasis}. This is a ubiquitous central pillar across all scales of biological organization, such as molecular circuits, physiological functions, and population dynamics. Failure of homeostatic control is associated with diseases including diabetes, autoimmunity, and obesity \cite{kotas2015homeostasis}. It is therefore vital to identify the regulated variables that the system aims to maintain at a stable setpoint.

Unlike simple human-made systems, where known pre-selected variables are under control, biological systems are characterized by multiple coupled control loops as well as other dynamic structures \cite{billman2020homeostasis}. A fundamental difference is that they are not divided to separate `plant' and `controller' entities, as is commonly assumed in control theory, but rather make up a complex network of interactions. In such a network regulated variables may be maintained at a stable setpoint, as in the classic example of the regulation of blood glucose concentration, a tightly regulated biological process in which the rates of glycolysis and gluconeogenesis are flexible variables \cite{kotas2015homeostasis}. A particularly interesting aspect is that, under certain conditions, biological homeostatic controllers may become oscillatory \cite{thorsen2014robust}. Indeed, while the occurrence of oscillations is generally avoided in control engineering, oscillatory behavior is ubiquitously found in natural systems \cite{krieger1982clocks, dunlap2004chronobiology, lloyd2012ultradian}. A well documented example is the Baroreflex control of the cardiovascular system where slow blood-pressure and heart-rate oscillations are observed \cite{di2009baroreflex}. The biological complexity makes it challenging to identify the regulated variables that the system actively maintains in the vicinity of an attractor and, as we further explain, dedicated algorithmic tools are required. 

Biological systems are commonly modeled by a set of dynamical equations where internal variables and control variables are not clearly separated. Therefore, the controlled objective and the control signal in a biological system model are implicit. Although control theory typically assumes the existence of a separate plant and controller (see Chapter 1.2 in \cite{aastrom2021feedback}), many theoretical results and analysis tools do not require such a separation \cite{cosentino2011feedback}. In what follows we elaborate on well known methods that are occupied with reconstructing control objectives from data and analyze their applicability for biological systems. We later introduce Identifying Regulation with Adversarial Surrogates (IRAS), a recently published algorithm that by directly addressing plant-controller coupling, is dedicated to control identification in biological systems. IRAS identifies variables that are regulated within narrow limits and is thus incompatible with biological systems which show sustained oscillations such as circadian rhythms \cite{moore1986physiology}. Last, we introduce Identifying Dynamic Regulation with Adversarial Surrogates (IDRAS), an extension of the IRAS algorithm, capable of identifying control objectives that are regulated with respect to a dynamic reference value. We test the algorithm on synthetic data of kinetic protein interactions, bacterial growth and division and on purely physical systems demonstrating excellent empirical results.

In Optimal Control and Reinforcement Learning a controller/agent, generically aims to optimize a value function, corresponding to an expected cumulative cost/reward function. In Inverse Optimal Control (IOC) and Inverse Reinforcement Learning (IRL), it is assumed that we observe trajectories of an optimal (or sometimes simply stable) controller or expert agent, which include both the states (of the system under control/the agent) and the control signals (actions). The goal of IOC/IRL is to infer the cost/reward function which the observed controller optimizes (see \cite{ab2020inverse, arora2021survey} for detailed surveys of both fields). In both IOC and IRL one has access to samples of a behaving system, acting according to some policy (usually a near-optimal one). These samples consist of both the system states and the external controls that drive the state-transitions. There is a clear separation between the states and the controls. Our biologically motivated setting corresponds to observing measurements of system variables without any prior knowledge of a separation between states and controls, as the control is implicit in the dynamical system and is not externally constructed. We are not aware of any IOC or IRL methods that deal with this type of problem.

The IRAS algorithm assumes that it is in general a \textit{combination} of the observed variables which is internally controlled. Therefore, given a set of measurements over time, $z(t) \in \mathbb{R}^m$, IRAS finds a combination $g(\cdot)$ of the measured variables, that is maintained around a setpoint,
\begin{equation}\label{eq:comb}
    g(z(t))\approx c_{\mathrm{set}},
\end{equation}
where we can assume $c_{\mathrm{set}}=0$ without loss of generality. In this work, we consider systems that are regulating some unknown combination of the observables over an unknown dynamically changing reference value,
\begin{equation}\label{eq:dynamic}
    g(z(t)) \approx c_e(t).
\end{equation}
By adding the subscript $e$ we emphasize that the behavior of $c_e(t)$ can be environment dependent. For example, circadian rhythms are regulated on molecular and physiological levels, but control parameters are entrained to the daylight cycle \cite{pittendrigh1964entrainment,duffy2005entrainment}. 
Common models for $c_e(t)$ are processes such as the Ornstein–Uhlenbeck process \cite{ricciardi1979ornstein} and oscillatory processes. Here we do not assume any specific form neither for the combination $g()$ nor for the process $c_e()$. For illustration, a general model that might yield observations that satisfy $g(z(t)) = c_e(t)$ (for a non-trivial $g()$-$c_e()$ pair) is,
\begin{equation}\label{eq:generalModel}
\begin{split}
    du_e &= f_{u_e}(u_e)dt + dW_{u_e},\\
    dx &= f_x(x,u_e)dt + dW_x,\\
    z_k &= h(x(t_k)) + \eta_{z_k},
\end{split}
\end{equation}
where $x \in \mathbb{R}^{n}$ is the state of the observed system, $u_e \in \mathbb{R}^{n_{u}}$ is an input from the environment, $z_k \in \mathbb{R}^{m}$ is the observation sampled at time $t_k$, $dW_{u_e}$ and $dW_x$ are Wiener processes and $\eta_{z_k}$ is a measurement noise. 

In general, any system identification procedure that does not decouple $g()$ from $c_e()$ will yield results that do not generalize to other environments where $c_{e'}(t)$ has a different behavior. It is therefore vital to separately identify the regulated combination from the dynamics of the reference. Specifically, the IRAS algorithm searches for a combination that is regulated about a fixed set-point, $g(z(t)) \equiv c_{\mathrm{set}}$ \cite{teichner2023identifying}. When applying it to a dynamic set-point it yields a negative answer - no regulation was found. In the next Section we present IDRAS, a purely data-driven algorithm that simultaneously learns a combination $g()$ and the dynamic process that it follows, $c_e()$.

\subsection{Related work}

Some dynamical systems obey conservation laws, where a combination of variables is constant along time trajectories of the system \cite{yang2019conservation}, as in \eqref{eq:comb}. Identifying conservation laws from observed data is an active research field, \cite{watters2017visual, santoro2017simple, hamrick2018relational, de2018end, chang2016compositional, tenenbaum2000global}, and it is important to note that, even given the differential equations of the system, identifying a conservation law analytically - or even proving its existence - is a difficult mathematical problem \cite{adem2012symmetry, lukashchuk2015conservation, adem2015conservation, el2016variational, el2018modulational, el2017new, el2017stability}. Methods for deriving the differential equations of a system \cite{brunton2016discovering} leave the question of identifying the conservation law unsolved. This problem is a sub-space of the one considered here, since the concept of conservation laws usually refers to well-defined dynamical systems with fixed parameter. Identifying biological regulation includes also the more general case where parameters are varying in time but still some combination is regulated around a set-point in the face of these perturbations.

Optimizing for conservation alone can lead to trivial quantities, such as predicting a constant $g(z)\!=\!c$ independent of $z$. In a recent paper, \cite{alet2021noether} refer to a non-trivial $g(\cdot)$ by the term \textit{useful conservation law}. To obtain a non-trivial solution, \cite{teichner2023identifying} define a measure of invariance, the ``Coefficient of Regulation" (CR) and an optimization algorithm that obtains meaningful invariants in \eqref{eq:comb}. Inferring meaningful dynamic regulatory processes, \eqref{eq:dynamic}, renders the task more difficult. Here we generalize IRAS to allow inferring meaningful dynamic regulatory process.

\section{Algorithm development}

\subsection{Problem formulation} 

We are interested in identifying empirically, from a set of measurements, a variable combination which tightly follows a dynamically changing reference value, where both are unknown. The combination could represent an internal quantity of high importance to the system, and the reference value could reflect temporal trends in the environment. Given observations $z_k \in {\mathbb{R}}^m$ at discrete times $k\in\{1,2,\ldots,N\}$, we search for $g_\theta:{\mathbb{R}}^{m} \rightarrow {\mathbb{R}}$, a function parameterized by $\theta$ such that,
\begin{equation}
    c_k = g_\theta(z_k),
\end{equation}
where the resulting time-series $c_k$ is the (learned) reference. We assume that this time-series follows some dynamics and it is therefore possible to learn a filter\footnote{Formally \eqref{eq:est} is a $1$-step predictor, but, following \cite{anderson2012optimal}, we refer to it as a filter.}, namely a predictor of the value of $c_k$ at time $k$ based on its previous values, formally given by
\begin{equation}\label{eq:est}
    \hat{c}_{k} = F_{\phi,\omega,\delta}(c_{k-T:k-1}).
\end{equation}
The structure of the learned filter $F$, and the meaning of its parameters, is detailed at the end of the present Section, and $c_{k-T:k-1} = [c_{k-T}, c_{k-T+1},\dots,c_{k-1}]$ with $T$ a hyper-parameter. The filtering error is,
\begin{equation}\label{eq:filtErr}
    e_k = c_k - \hat{c}_{k} \overset{\Delta}{=} \mathcal{E}_\Omega(z_{k-T:k}),
\end{equation}
and is a function of the parameters $\Omega = [\theta, \phi, \omega, \delta]$. Our goal is to learn both $g_\theta()$ and $F_{\phi,\omega,\delta}()$ such that the error is small, namely $e_k \approx 0$ for all $k$. We note that a straightforward optimization yields the trivial pair, $g() \equiv 0$, $F() \equiv 0$.
\subsection{Using adversarial surrogates to identify regulation
}
The problem formulated is in fact a generalization of problem \eqref{eq:comb}. It is immediate to verify that by setting $F_{\phi,\omega,\delta}(\cdot) \equiv 0$, we obtain $\mathcal{E}_\Omega(z_{k-T:k}) = g_\theta(z_k)$,\\$e_k = c_k$ and thus we optimize to find a combination $g_\theta(\cdot)$ such that $c_k \approx 0$, namely regulated to a narrow range around a fixed value.

This simpler problem was addressed in \cite{teichner2023identifying} by utilizing comparisons between original time-series and random (though possibly constrained) shuffled time-series. A regulated combination is presumably composed of components that co-vary to compensate and buffer perturbations.  Therefore, shuffling the temporal order of each component separately would ruin these co-variations and greatly increase the variance of the combination over time. To quantify this notion, a ``coefficient of regulation" is defined as the ratio between the standard deviations of the combination in the original data $c_k$ and in the shuffled data $\tilde{c_k}$. In \cite{teichner2023identifying} it was found that straightforward optimization of this measure is insufficient to escape trivial solutions and artefacts. Rather, a two-player algorithm was constructed which alternates between minimizing the CR and modifying the shuffled time-series. This algorithm terminates when the CR can no longer be minimized, while at the same time reproducing general geometric features of the data in parameter space.

\paragraph{IRAS algorithm (recap)}
Given observations $z_k \in {\mathbb{R}}^m$ IRAS identifies a combination $g_\theta:{\mathbb{R}}^{m} \rightarrow {\mathbb{R}}$, for which $c_k=g_\theta(z_k) \approx c_{\mathrm{set}}$ for all $k$. The iterative algorithm alternates between minimizing the CR,
%
    $\nicefrac{\sigma(c_{1:N})}{\sigma(\tilde{c}_{1:N})}$
%
($\sigma(\cdot)$ denotes the standard-deviation), and modifying the surrogate time-series $\tilde{c}_k$ by minimizing the distribution distance $\mathcal{D}(c_{1:N}, \tilde{c}_{1:N})$ under an information constraint.

Relying on similar concepts, the next Section introduces the generalization of this algorithm to time-varying regulation set-points.

\subsection{Identifying Dynamic Regulation with Adversarial Surrogates}\label{sec:IDRAS}

IDRAS is an iterative algorithm consisting of two competing players, a generalization of the IRAS algorithm. The first player aims to minimize the CR, a measure of invariance (see \cite{teichner2023identifying} Section 1), while the second player aims to render the task of the first player more difficult by forcing it to extract information about the temporal structure of the data, which is absent from time-shuffled “surrogate” data. 

\paragraph{Combination player}At iteration $i$, the first player, realized by artificial neural networks, sets the parameters $\Omega$ to minimize the CR,
\begin{equation}\label{eq:combplayer}
\begin{split}
    \Omega^{(i)} = \argmin_{\Omega} \frac{\sigma(e_{T+1:N})}{\sigma(\tilde{e}_{T+1:N}^{(i)})} , 
\end{split}
\end{equation}
where $\sigma(e_{T+1:N})$ and $\sigma(\tilde{e}_{T+1:N}^{(i)})$ are the standard-deviations of the time-series $e_k$ and $\tilde{e}_k^{(i)}$ respectively. The time-series $e_k$ is the filtering error on the original data and the time-series $\tilde{e}^{(i)}_k$ is the filtering error on surrogate data, the outcome of a resampling procedure using a resampling function $\zeta^{(i-1)}()$ that was set by the second player in the previous iteration,
\begin{align}\label{eq:tilde}
        e_k &= \mathcal{E}_\Omega(z_{k-T:k}),\tilde{e}_k^{(i)} = \mathcal{E}_\Omega(z_{k-T:k-1},\tilde{z}_k^{(i)}), \nonumber\\
        \tilde{z}_k^{(i)} &\sim P^{\zeta^{(i-1)}}_{\tilde{z}}(\tilde{z}_k^{(i)} \mid z_{k-T:k-1})
        =
        P_z(\tilde{z}_k^{(i)})\zeta^{(i-1)}(\tilde{e}_k^{(i)}),
\end{align}
where $P_z(z)$ is the probability of observing a measurement $z$ at some random time. 

Optimizing \eqref{eq:combplayer}, the combination player searches for a combination $g_\theta(\cdot)$, whose filtering error (which is the $1$-step prediction error) $\sigma(e_{T+1:N})$ is small w.r.t.~the error in predicting a sample from a random point in time, $\sigma(\tilde{e}^{(i)}_{T+1:N})$. This encourages the combination player to find a useful $(g,F)$ pair, representing a meaningful underlying quantity in the observed system. 

We note that for $\zeta(\cdot) \equiv 1$ the time-series $\tilde{z}_k$ is a na\"ive random permutation of the time-series $z_k$. Optimizing \eqref{eq:combplayer} w.r.t.~such unconstrained permutation leads to artifacts. To illustrate consider that one of the observables is merely a Wiener process. The variance of a $1$-step prediction error is proportional to $dt$ while the variance of predicting a random sample increases with $N$, the length of the time-series. Therefore, without constraining the permutation, the CR of a trivial combination that outputs the Wiener process approaches zero as $N$ increases. This leads to the identification of the Wiener process as the control objective, clearly an artifact. We refer the reader to Section 1.B in \cite{teichner2023identifying} for a proof and a detailed explanation. To avoid these artifacts the second player constrains the time-series $\tilde{z}_k$ by setting the resampling function $\zeta(\cdot)$. 

\paragraph{Shuffle player}The second player at iteration $i$ makes use of the current proposed solution of the combination player, $\Omega^{(i)}$, to create a new shuffled time-series $\tilde{z}^{(i+1)}_k$, which better resembles the statistical structure of the data under the $1D$ projection $\mathcal{E}_{\Omega^{(i)}}$. Formally, this corresponds to the selection of a resampling function $\zeta^{(i)}(\cdot)$ that minimizes the distributional distance,
\begin{equation}\label{eq:optShuffle}
    \begin{split}
        \zeta^{(i)} = \argmin_{\zeta} \mathcal{D}(e_{T+1:N}, \tilde{e}_{T+1:N}),
    \end{split}
\end{equation}
where
\begin{align}\label{eq:tildeS}
        e_k &= \mathcal{E}_{\Omega^{(i)}}(z_{k-T:k}), \tilde{e}_k = \mathcal{E}_{\Omega^{(i)}}(z_{k-T:k-1},\tilde{z}_k),\nonumber\\
        %
        %
        \tilde{z}_k &\sim P^{\zeta}_{\tilde{z}}(\tilde{z}_k \mid z_{k-T:k-1}) = P_z(\tilde{z}_k)\zeta(\tilde{e}_k).
\end{align}
\begin{Lemma}\label{lem:shuffleOpt}
(Shuffle player's optimal solution.) The shuffle player, who only has access to the error time-series, can solve optimization \eqref{eq:optShuffle} and obtain $\mathcal{D}=0$ by choosing
\begin{equation}\label{eq:zetaSol}
    \zeta\left(\mathcal{E}_{\Omega^{(i)}}(\cdot)\right) = \frac{P_{z_{(T)}}(\cdot)}{P_{z^*_{(T)}}(\cdot)},
\end{equation}
where $P_{z_{(T)}}(z_{k-T:k})$ is the probability of observing the sequence $z_{k-T:k}$ (at a random time $k$) and 
\begin{equation}\label{eq:tsShuffle}
    \begin{split}
        P_{z^*_{(T)}}(z_{k-T:k-1},z) &= P_{z_{(T-1)}}(z_{k-T:k-1})P_z(z)
    \end{split}
\end{equation}
is the probability of observing the sequence $z_{k-T:k-1}$ followed by a time-random observation, sampled from the na\"ivly permuted time-series. 
\end{Lemma}
\paragraph{Proof of Lemma \ref{lem:shuffleOpt}.}In what follows we prove that the resampling function defined in \eqref{eq:zetaSol} yields $\mathcal{D}=0$ in \eqref{eq:optShuffle}. Based on \eqref{eq:tildeS} define 
\begin{equation*}
    \begin{split}
        &P_{\tilde{z}_{(T)}}(z_{k-T:k-1},\tilde{z}_k) = P_{z_{(T-1)}}(z_{k-T:k-1})P^\zeta_{\tilde{z}}(\tilde{z}_k \mid z_{k-T:k-1}),
    \end{split}
\end{equation*}
the probability that the shuffle player will concatenate $\tilde{z}_k$ to the observed sequence $z_{k-T:k-1}$. In \eqref{eq:optShuffle}-\eqref{eq:tildeS}, the time-series $e_k$ inherits its distribution via the $1D$ projection from the distribution $P_{z_{(T)}}$ and the time-series $\tilde{e}_k$ inherits its distribution from $P_{\tilde{z}_{(T)}}$. We will show that for the choice of the resampling function in \eqref{eq:zetaSol}, the two $1D$ projections coincide. Note that
%
    \begin{align}\label{eq:P_tilde_z}
        &P_{\tilde{z}_{(T)}}(z_{k-T:k-1},\tilde{z}_k) 
        = P_{z_{(T-1)}}(z_{k-T:k-1})P^\zeta_{\tilde{z}}(\tilde{z}_k \mid z_{k-T:k-1})\nonumber\\
        &\overset{(i)}{=} P_{z_{(T-1)}}(z_{k-T:k-1})P_z(\tilde{z}_k)\zeta(\tilde{e}_k)\nonumber\\
        &\overset{(ii)}{=} P_{z^*_{(T)}}(z_{k-T:k-1},\tilde{z}_k)\zeta(\tilde{e}_k),
    \end{align}
%
where $(i)$ is by the definition of $P_{\tilde{z}}^{\zeta}$ in \eqref{eq:tildeS} and $(ii)$ by the definition of $P_{z^*(T)}$ in \eqref{eq:tsShuffle}. The derivation in \eqref{eq:P_tilde_z} implies that
\begin{equation*}
    P_{\tilde{z}_{(T)}}(\tilde{e}_k=\mathcal{E}_{\Omega^{(i)}}(z_{k-T:k-1},\tilde{z}_k)) = P_{z^*_{(T)}}(\tilde{e}_k)\zeta(\tilde{e}_k),
\end{equation*}
from which by substituting \eqref{eq:zetaSol}, $\zeta(\tilde{e}_k)=\frac{P_{z_{(T)}}(\tilde{e}_k)}{P_{z^*_{(T)}}(\tilde{e}_k)}$, it directly follows that $P_{\tilde{z}_{(T)}}(\tilde{e}_k)=P_{z_{(T)}}(\tilde{e}_k)\qquad\qquad\square$.
%
%
\begin{figure}
\centering
\includegraphics[width=0.7\textwidth]{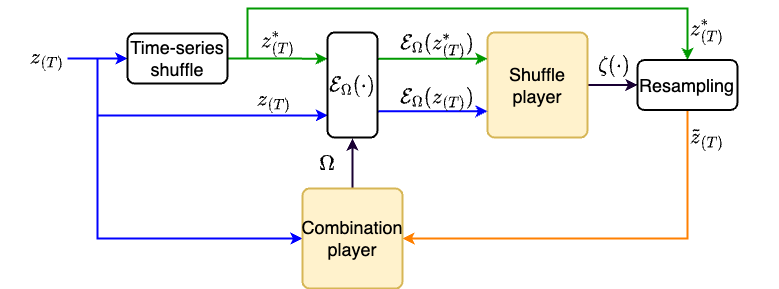}
\caption{IDRAS algorithm outline. The observation time-series $z$ is permuted according to \eqref{eq:tsShuffle} to create the unconstrained series $z^*$. The shuffle player, only exposed to the $1D$ projections under $\mathcal{E}_\Omega()$, sets the resampling function $\zeta()$ used to resample $\tilde{z}$ from $z^*$ such that the distributions under the projection are identical, \eqref{eq:optShuffle}. Then the combination player, given $z$ and $\tilde{z}$, updates the parameters $\Omega$ towards minimizing the CR, \eqref{eq:combplayer}. These steps continue to iterate until no further improvement is possible. The block $\mathcal{E}_\Omega$, based on \eqref{eq:filtErr} and whose architecture is detailed in Figure \ref{fig:GammaBlock}, replaces the function $g()$ in Fig.~3 in \cite{teichner2022enhancing}. \label{fig:SimpleFlowIDRAS}}
\end{figure}

The two players mutually inform each other of their current step results, and the process continues iteratively until the combination player can no longer decrease the CR in \eqref{eq:combplayer}. We refer to this algorithm as IDRAS, depict its outline in Figure \ref{fig:SimpleFlowIDRAS} and pseudo-code in Algorithm \ref{alg:cap}. We note that Algorithm \ref{alg:cap} supports multiple observed systems by calculating the CR and $\zeta()$ per system.
%
\begin{algorithm}[t!]
\caption{IDRAS}\label{alg:cap}
\begin{algorithmic}[1]
\STATE \textbf{Input:} $\{z^{(s)}_{1:N_s}\}_{s=1}^S$, $T$, $\eta$, $\texttt{nEpochs}$
\STATE \textbf{Initialize:} $\zeta^{(s)} \equiv 1 \forall s \in [1,S]$,$\Omega \sim \mathcal{N}(0,1)$ \COMMENT{i.i.d}
\FOR{$i=1,\dots, \texttt{nEpochs}$}
\FOR{$s=1,\dots,S$}
\FOR{$k=T+1,\dots,N_s$}
\STATE $c_k=g_{\theta}(z^{(s)}_k)$, $\hat{c}_{k}=F_{\phi,\omega,\delta}(c_{k-T:k-1})$
\STATE $e_k=c_k-\hat{c}_{k}$, $\tilde{z}_k \sim P_{\tilde{z}}^{\zeta^{(s)}}$ \COMMENT{see \eqref{eq:tilde}}
\STATE $\tilde{c}_k=g_{\theta}(\tilde{z}_k)$, $\tilde{e}_k=\tilde{c}_k-\hat{c}_{k}$
\ENDFOR
\STATE $\mathrm{CR}^{(s)}=\frac{\sigma(e_{T+1:N})}{\sigma(\tilde{e}_{T+1:N})}$ \COMMENT{combination player, \eqref{eq:combplayer}}
\STATE $\zeta^{(s)}\left(\mathcal{E}_\Omega(\cdot)\right) = \frac{P_{z^{(s)}_{(T)}}(\cdot)}{P_{z^{(s)*}_{(T)}}(\cdot)}$ \COMMENT{shuffle player,\eqref{eq:zetaSol}}
\ENDFOR
\STATE $\Omega \gets \Omega - \eta \nabla_{\Omega}\operatorname{E}_s\left[\mathrm{CR}^{(s)}\right]\quad$\COMMENT{Update step}
\ENDFOR
\STATE \textbf{return} $\Omega$
\end{algorithmic}
\end{algorithm}
\paragraph{Filter}The architecture of the filter within the block $\mathcal{E}_\Omega()$ in Fig.~\ref{fig:SimpleFlowIDRAS} has many degrees-of-freedom and can be chosen by the user according to prior knowledge regarding the nature of the dynamic reference. To impose few constrains on the filter $F(\cdot)$ it can be implemented by a fully connected deep neural-network. 

\begin{figure}
\centering
\includegraphics[width=0.5\textwidth]{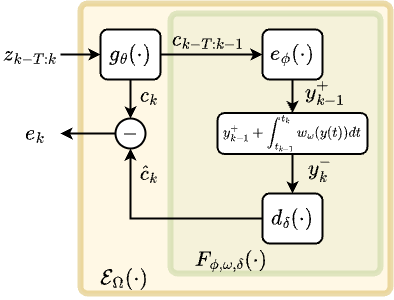}
\caption{Architecture of the $\mathcal{E}_\Omega(\cdot)$ block \eqref{eq:filtErr} from Fig.~\ref{fig:SimpleFlowIDRAS}. The filtering error $e_k$ is the difference between the learned reference value $c_k$ and its $1$-step prediction $\hat{c}_k$. The filter $F_{\phi,\omega,\delta}(\cdot)$, \eqref{eq:filterBlock}, infers a latent-state $y^+_{k-1}$ and time-advances it to the state $y^-_k$ from which the estimation $\hat{c}_k$ is decoded.\label{fig:GammaBlock}}
\end{figure}

Dealing with biological systems, we assume that the dynamics of the reference can be modeled by a continuous-time, time-invariant latent model (see \cite{li2020scalable,kidger2021neuralsde} for details on integrating differential equations using neural-networks). Our filter-block contains three parts: \emph{(i)} An encoder $e_\phi: \mathbb{R}^{T} \rightarrow \mathbb{R}^{n_y}$, that given $T$ consecutive values of the reference, $c_{k-T:k-1}$, infers a latent state $y^+_{k-1} \in \mathbb{R}^{n_y}$ (with $n_y$ a user-defined hyper-parameter), \emph{(ii)} A drift function $w_\omega: \mathbb{R}^{n_y} \rightarrow \mathbb{R}^{n_y}$ describing the deterministic term in the dynamics of the latent state that serves to time-advance the latent state $y^+_{k-1}$ to $y^-_{k}$, \emph{(iii)} An emission function $d_\delta: \mathbb{R}^{n_y} \rightarrow \mathbb{R}$ that decodes the $1$-step predicted value $\hat{c}_{k}$ from the latent state $y^-_{k}$. The following set of equations describe the filter-block \eqref{eq:est},
\begin{equation}\label{eq:filterBlock}
\begin{split}
    y^+_{k-1} &= e_\phi(c_{k-T:k-1}),\\
    y^-_{k} &= y^+_{k-1} + \int_{t_{k-1}}^{t_k}w_\omega(y(t))dt,\\ 
    \hat{c}_{k} &= d_\delta(y^-_{k}),
\end{split}
\end{equation}
depicted, as part of the $\mathcal{E}_\Omega()$ block in Figure \ref{fig:GammaBlock}.
\subsection{Performance assessment metric}\label{sec:performance}
IDRAS, an unsupervised learning algorithm, yields a time-series of a regulated quantity, $c_k$, and the corresponding time-series of $1$-step predictions, $\hat{c}_{k}$. To assess whether a quantity that follows a dynamic reference value was indeed found, we calculate the normalized prediction error energy $\rho(c, \hat{c})$. 
\begin{definition}\label{def:rho}
The normalized error energy for two time-series $x_{1:N}$ and $\hat{x}_{1:N}$ is defined,
\begin{align}
        \rho(x_{1:N},\hat{x}_{1:N}) &= \frac{\sum_{k=1}^N (\bar{x}_k - \bar{\hat{x}}_k)^2}{\sum_{k=1}^N (\bar{x}_k)^2},\quad
        \bar{x}_k = \frac{x_k-E[x]}{\sigma(x)},\nonumber
\end{align}
where $\operatorname{E}[x]=\frac{1}{N}\sum_{k=1}^N x_k$ and $\sigma^2(x)=\frac{1}{N}\sum_{k=1}^N(x_k-\operatorname{E}[x])^2$, and similary for $\hat{x}$.
\end{definition}
The normalized error energy $\rho(c,\hat{c})$ allows for an immediate assessment of the significance of the identified combination. A low $\rho$ value ($\rho < 0.5$) indicates that a combination that follows a dynamic reference was found. 
We remind the reader that the shuffle player in IDRAS guarantees that the algorithm won't converge to trivial combinations such as $c_k\equiv \mathrm{const}$.

%
\section{Validation}\label{sec:validation}
After presenting the construction of IDRAS, we seek to validate it on datasets with a known control objective, so that the quality of the results can be assessed. We chose two validation examples of biological models: a kinetic model of protein interactions and a model of bacterial life cycle. Additionally, to demonstrate the efficiency of our algorithm in studying physical systems, we validate the algorithm on a dataset that serves for benchmarking machine-learning algorithms. To assess the performance, we calculate the normalized error energy between the known control objective $c^*_k$ and the output of IDRAS $c_k$ and consider $\rho(c,c^*) < 0.1$ as an excellent empirical result. The architecture of all parameterized functions are listed in SI Appendix, section 1.1 and code reproducing all examples is available at \href{https://github.com/RonTeichner/IRAS}{https://github.com/RonTeichner/IRAS}.
\subsection{A kinetic model of interactions}
We first validate IDRAS on simulated data generated from a kinetic model that describes regulatory interactions in the production of two proteins incorporating a feedback loop. In the considered model (inspired by \cite{el2021biological}), the total amount of two proteins $P$ and $S$, namely $P+S$, is controlled by $M$, the mRNA molecule that is produced through a process of constitutive transcription at a rate $K$ and degraded with first order kinetics at a rate $\gamma_M$. The model is under constant perturbations to the protein expression rate $K$. These might be caused by ambient temperature dynamics that were found to have large passive effects on both mRNA synthesis and decay rates \cite{sidaway2014direct}. The model is described by the differential equations  

\begin{equation}\label{eq:kineticModel}
\begin{aligned}
    K(t)&= K_0(1+0.5\cos(2\pi\frac{1}{\tau_K} t + \phi_K)),\\
    d{M}&= (K(t)-f(P+S)-\gamma_M M)dt,\\
    d{P}&= (k_P M-\gamma_P P)dt + \eta_P dW_P,\\
    d{S}&= (k_S M-\gamma_S S)dt + \eta_S dW_S,
\end{aligned}
\end{equation}

where the mRNA $M$ and the two proteins $P,S$ are linked in a feedback loop. Both $P$ and $S$ are positively affected by $M$, with their steady-state values proportional to it. The concentration $M$ in turn, is negatively affected by the sum $P+S$, with the strength of this negative feedback given by the rate constant $f$. We note that \eqref{eq:kineticModel} is a model of dimension $5$, comprising $3$ state variables and an oscillating input $K$ that results from an underlying state-space model of dimension at least $2$ (a $1D$ model cannot produce oscillations).

\begin{figure}[t!]
\centering
\includegraphics[width=0.6\textwidth]{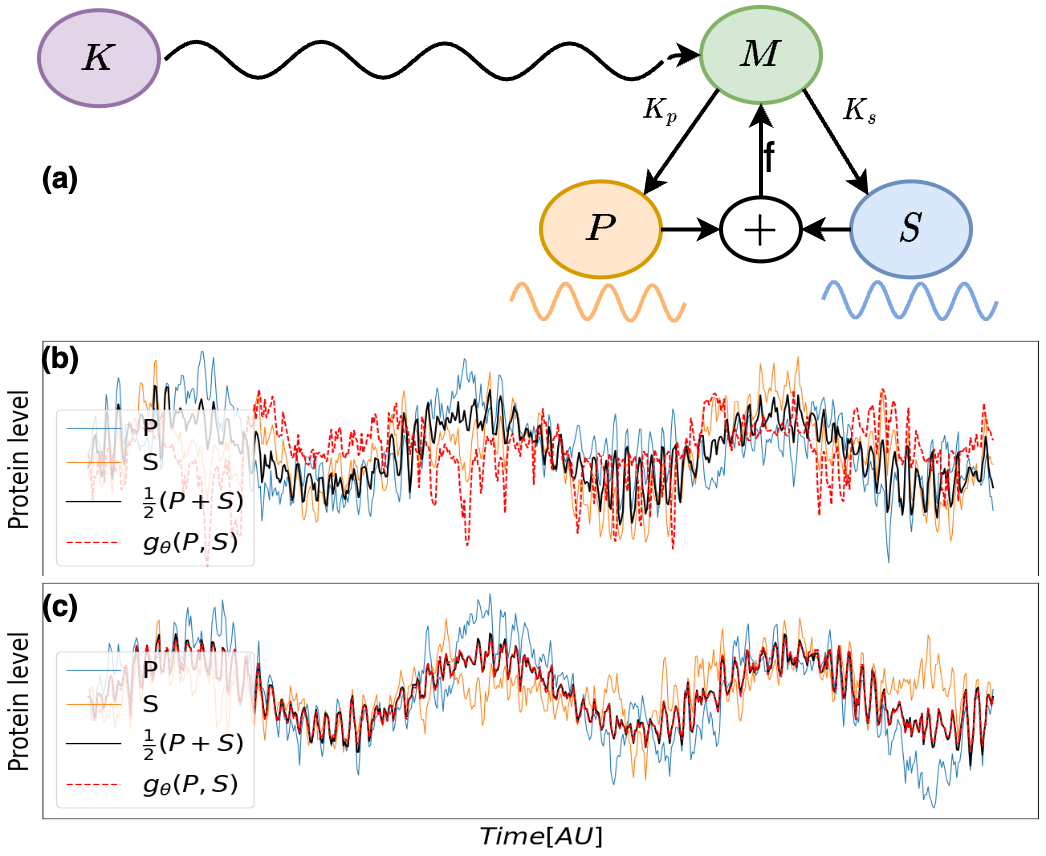}
\caption{(a) Illustration of the closed-loop model \eqref{eq:kineticModel}. The mRNA molecule $M$ induces the production of proteins $S$ and $P$ and receives a negative feedback of their sum. (b) The IRAS algorithm does not converge, due to the lack of a regulated constant combination. The combination $P+S$ (black) along with the output of IRAS (dashed red). (c) IDRAS algorithm captures the control objective and its oscillating trend.\label{fig:Proteins}}
\end{figure}

Small changes in $S$ or $P$, modeled by the an increments of the Wiener processes $dW_P$ and $dW_S$, induce swift and sharp changes in the transcription of $M$ and maintain $P\!+\!S$ around a reference level
\begin{equation}
    c^*(t) \overset{\Delta}{=} (P+S)(t) = \frac{1}{f+\cfrac{\gamma_M\gamma_P\gamma_S}{k_P\gamma_S+k_S\gamma_P}} K(t),    
\end{equation}
which is reflected in a high negative correlation between $S$ and $P$ \cite{sidaway2014direct}. In \eqref{eq:kineticModel} we identify the components of the general model \eqref{eq:generalModel}: $u_e = [K,M]$, $x = [P,S]$. Our observations contain the levels of the two proteins $P$ and $S$ (an observation taken in labs by measuring fluorescence intensity of constitutively expressed proteins \cite{CHOI20231}), $z_k = [P(t=\nicefrac{k}{f_s}),S(t=\nicefrac{k}{f_s})]$, where $k\in[1,\dots,N]$ and $f_s$ is the sampling rate. Figure \ref{fig:Proteins}a illustrates the kinetic interactions model and SI Appendix section 1.2 lists the parameter values. 

We ran both IRAS and IDRAS algorithms in search of the control objective. Figure \ref{fig:Proteins}b depicts the output (dashed red) of $g_\theta(P,S)$ trained by the IRAS algorithm. Due to the lack of a regulated constant combination, IRAS did not converge and scored $\rho_{\mathrm{IRAS}}(c,c^*)=1.82$ (see Definition \ref{def:rho}). 

When running IDRAS, we have a simple measure to assess whether a quantity that follows a dynamic reference value was found, the measure of the normalized prediction error energy defined in Section \ref{sec:performance}. A score of $\rho_{\mathrm{IDRAS}}(c,\hat{c})=0.148$ was obtained, indicating that a combination that tightly follows a dynamic reference was found. The validation score is $\rho_{\mathrm{IDRAS}}(c,c^*)=0.012$, indicating that the combination $P+S$ was precisely found, despite its oscillating nature, as depicted in Figure \ref{fig:Proteins}c.
\subsection{Bacterial life cycle}
The next example we consider is a realistic biological model of bacterial life cycle. We shall focus here on bacterial growth homeostasis, where growth and division proceed for many generations with significant variability and statistical stability. This is a problem with a long history and on which a large body of data is available. We apply our algorithm to simulation data, which mimic experimental measurements but where the regulation is known, and show that IDRAS detects the correct mode of regulation. 

Most bacteria grow smoothly and divide abruptly, consistent with a threshold crossing by some division indicator at the single-cycle timescale \cite{amir2014cell,grilli2017relevant,susman2018individuality}; the threshold itself can have dynamics over multiple cycles. Three types of division indicators corresponding to different regulation modes have been proposed: cell size (“sizer” control mechanism), added size (“adder” mechanism) and elapsed time (“timer”) \cite{jun2015cell, ho2018modeling}. In a recent paper, Luo et al. \cite{luo2023stochastic} demonstrate that commonly used heuristic tools uncover the correct mode of regulation only under very restricted conditions. We reproduce these results and show that in contrast to the commonly used tools, IDRAS accurately identifies the cell-division mechanism. 

\begin{figure}[t!]
\centering
\includegraphics[width=0.6\textwidth]{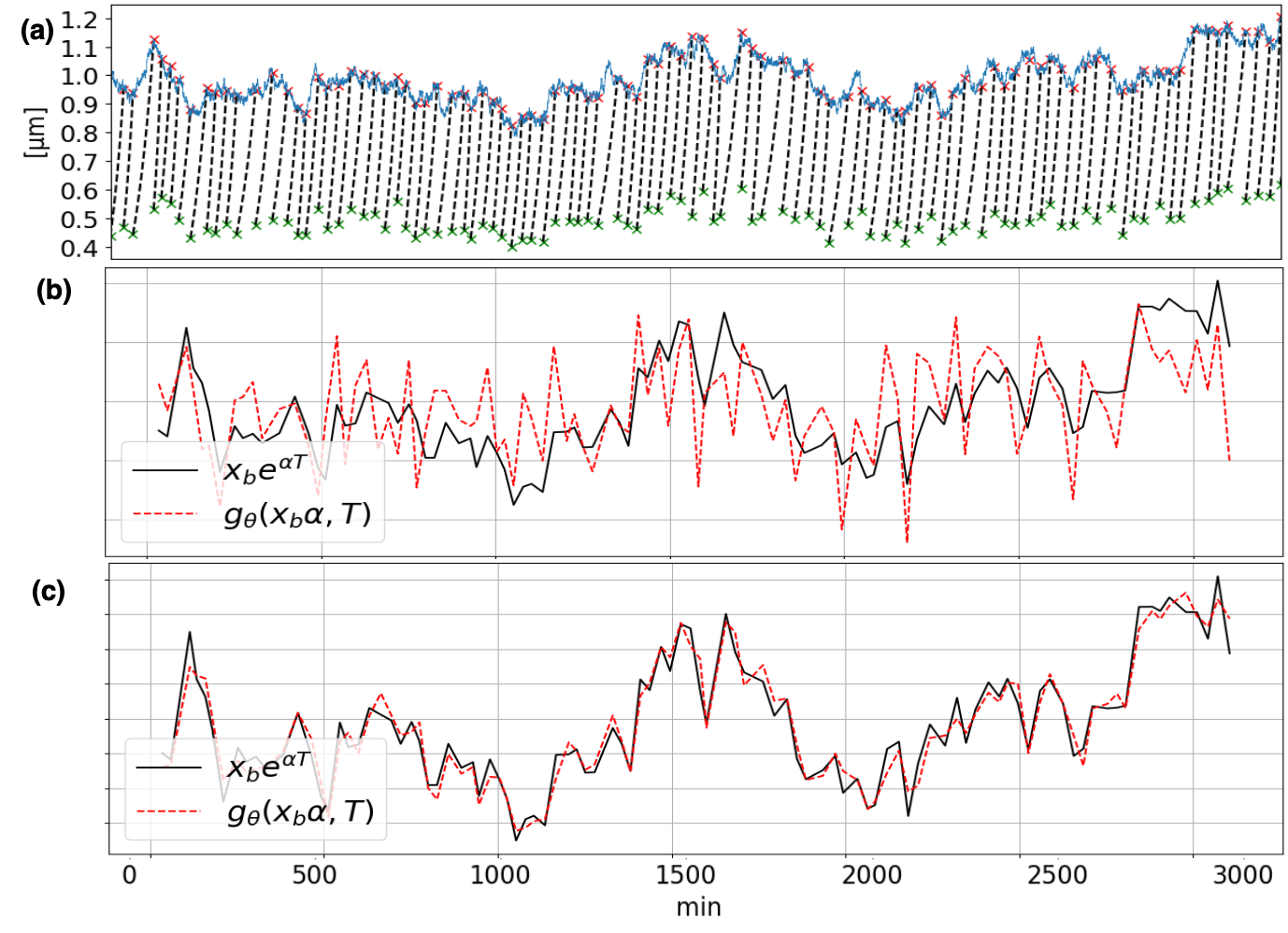}
\caption{(a) Bacteria life-cycles. The blue curve is the stochastic threshold process $u(t)$, \eqref{eq:OU}, and the black curve is the cell-size over time, \eqref{eq:cellSize}, such that the cell divides once it reached the dynamic threshold.
(b) IRAS does not decouple the cell-size mechanism (black-curve) from the threshold trend $c(t)$ and yields a combination (dashed-red) that represents their mixture. (c) IDRAS captures the division-mechanism
.  \label{fig:Bacteria}}
\end{figure}

The inter-generation threshold $u(t)$ is modeled by a stochastic Ornstein-Uhlenbeck process,
\begin{equation}\label{eq:OU}
du = \frac{\mu_u-u}{\tau_u}dt + \sqrt{2\frac{\sigma^2_u}{\tau_u}}dW
\end{equation}
where the term $dW$ is an increment of a Wiener process \cite{luo2023stochastic}. On the single-cycle scale, the $k$\textsuperscript{th} cell grows exponentially, 
\begin{equation}\label{eq:cellSize}
\begin{split}
x(t) &= x_{k,b} e^{\alpha_k (t-t_{k,b})}, \quad t_{k,b} \leq t \leq t_{k,d}\\
x_{k,b} &= x(t_{k-1,d})f_k,\qquad t_{k,b} = t_{k-1,d}\\
t_{k,d} &= \mathrm{argmin}_t\{t \geq t_{k,b} \mid x_{k,b} e^{\alpha (t-t_{k,b})} = u(t)\},
\end{split}
\end{equation}
where $x_{k,b}\overset{\Delta}{=}x(t_{k,b})$ is the birth size, $x_{k,d}\overset{\Delta}{=}x(t_{k,d})$ is the size at division, $\alpha_k$ is the exponential growth rate and $f_k$ is the division fraction. Here we simulate a \textit{sizer} mechanism such that the cell divides when its size crosses the threshold $u(t)$. Figure \ref{fig:Bacteria}a depicts a simulated lineage over time - the cell-size, $x(t)$ (dashed-black), and the stochastic threshold $u(t)$ (blue). SI Appendix section 1.3 lists the parameter values.

In \eqref{eq:OU} and \eqref{eq:cellSize} we identify the components of the general model, \eqref{eq:generalModel}, with $u_e=u$ and $x=x$. Our observations, derived from $x(t)$, contain the initial size, growth rate and the cycle duration, $z_k=[x_{k,b},\alpha_k,T_k=t_{k,d}-t_{k,b}]$ where $\alpha_k=\frac{1}{T_k}\log(\frac{xd}{xb})$. We note that our choice of the feature vector $z_k$ renders IDRAS's task harder as now to correctly detect the sizer mechanism the network has to learn the combination $g()=x_b e^{\alpha T}$ and not just $g()=x_d$, in case $x_d$ was an entry of $z_k$.

We ran both IRAS and IDRAS algorithms in search of the division mechanism. We chose a realistic parameter set for which heuristic identification methods based on data correlations fail to detect the control mechanism correctly \cite{luo2023stochastic}.
Figure \ref{fig:Bacteria}b depicts the output (dashed red) of $g_\theta(x_b,\alpha,T)$ trained by the IRAS algorithm vs the ground-truth output, the sizer division mechanism $x_b e^{\alpha T}$. Here IRAS, optimizing objective \eqref{eq:comb}, outputs a combination that results from fusing together the division mechanism and the dynamic threshold $u(t)$ (with $\rho_{IRAS}(c,c^*)=0.66$), as we now explain. In our model, the timescale of the threshold is significantly slower than the time scale of single-cells, $\tau_u \gg T$, such that the threshold is approximately constant along a single-cell cycle, $u(t) \approx u_k$ for $t_{k,b} \leq t \leq t_{k,d}$. The birth size, about half the size at division of the previous cell is thus $x_{k,b} \approx 0.5u_k$, and the size at division is (by definition) $x_{k,d} \approx u_k$. The single-cell cycle regulated combination is therefore,
\begin{equation}
    x_{k,b}e^{\alpha_k T_k} - u_k \approx x_{k,b}e^{\alpha_k T_k} - 2x_{k,b} \approx 0, 
\end{equation}
a mixture of two terms. The first, $x_{k,b}e^{\alpha_k T_k}$, represents the sizer mechanism while the second, $-2x_{k,b}$, is an influence of the dynamic threshold process. 

IDRAS, optimizing objective \eqref{eq:dynamic}, decouples the two by separately learning the threshold dynamics and a combination which is regulated w.r.t.~this dynamic reference value. Figure \ref{fig:Bacteria}c depicts the precise identification of the sizer mechanism with a validation score $\rho_{IDRAS}(c,c^*)=0.046$. The normalized prediction error energy is $\rho_{IDRAS}(c,\hat{c})=0.23$, fitting the expected value derived from the noise term in \eqref{eq:OU} up to $8\%$ ($\nicefrac{2\bar{T}}{\tau_c}$, where $\bar{T}$ is the mean cell-cycle, about $25$ minutes).
\subsection{Identifying Complex Physical Equations}
To further challenge the IDRAS algorithm, we evaluate it on a broad range of physics problems, taken from the seminal Feynman Lectures on Physics \cite{feynman1965feynman}, also used in the recently published “Feynman Symbolic Regression Database” (FSReD) \cite{udrescu2020ai}. We simulate physical equations incorporating oscillating terms and expect IDRAS to decouple the oscillating terms from the quantities they follow. 

One such equation is equation $II.6.15b$, taken from Volume 2 in \cite{feynman1965feynman}, which describes the electric field induced by a dipole. 
The dipole, with a dipole moment $P_d$ (the product of the charges and their separation) induces, at distance $r$ and angle $\omega t$, an electric field whose transverse component is $E_f$,
\begin{equation*}
    E_f = \frac{3}{4\pi \epsilon} \frac{P_d}{r^3} \cos(\omega t) \sin(\omega t).
\end{equation*}
We simulate $100$ time-series of length $100$ in which the dipole spins with angular velocity $\omega$. For each sample, the observables ($E_f,\epsilon,P_d$ and $r$) are sampled uniformly in $[1,5]$ (although not physically valid, it is the way the FSReD dataset is synthesized \cite{udrescu2020ai}). We expect IDRAS to identify the combination
\begin{equation*}
    g_{\theta^*}(E_f,\epsilon,P_d,r)=E_f \frac{4\pi \epsilon r^3}{3P_d} = \cos(\omega t) \sin(\omega t) = c^*(t).
\end{equation*}
To quantify the performance of IDRAS, we compute $\rho(c,c^*)$ for each of three chosen examples from the fields of electric-field forces ($I.12.11$), non-linear responses ($I.50.26$), and electric dipoles $II.6.15b$. As shown in table \ref{sample-table}, the agreement is excellent.
\begin{table}[h!]
\caption{IDRAS captures physical relations} \label{sample-table}
\begin{center}
\begin{tabular}{lll}
&Equation  &$\rho(c,c^*)$ \\
\hline \\
$I.12.11$         &$F = q*(E_f + Bv\sin(\omega t))$ & 0.007 \\
$I.50.26$         &$x = x_1(\cos(\omega t) + \alpha \cos^2(\omega t))$ & 0.022\\
$II.6.15b$        &$E_f = \frac{3}{4\pi \epsilon} \frac{P_d}{r^3} \cos(\omega t) \sin(\omega t$ & 0.059 \\
\end{tabular}
\end{center}
\end{table}

\section{Discussion}

Detecting regulatory processes in dynamic data is a technically challenging problem with many potential applications. Recently an algorithm named IRAS was introduced \cite{teichner2023identifying} that receives as input raw dynamic measurements and provides combinations of the observables that are maximally conserved across time. Here we presented IDRAS, a generalization of IRAS, capable of identifying control objectives that are regulated w.r.t.~a dynamical reference value. This algorithm allows the identification of control objectives in biological systems which show sustained oscillations, such as circadian rhythms.

The algorithm searches for a combination of the observables that tightly follows a dynamic reference value by optimizing a quantitative measure — the Coefficient of Regulation (CR) \cite{teichner2023identifying}. In IDRAS, the CR characterizes the sensitivity of the $1$-step prediction error of a combination to destroying the temporal order of the observed time-series. IDRAS iterates between two players, one minimizing the CR and the other creating successively more constrained shuffled ensembles, and converges when the two players cannot further improve. The algorithm then outputs a coupled pair - the combination and the dynamics it follows. This allows for an immediate assessment of the significance of the identified combination by calculating the normalized prediction error (see Section \ref{sec:performance}). We provide validation in three distinct realistic examples demonstrating excellent empirical results.

Biological systems are often ``reversed engineered" to build mathematical models based on observed data \cite{ljung1998system,schmidt2009distilling,daniels2015automated,brunton2016discovering,shen2021finding,haber2022learning,chen2022automated}. And yet, identifying control objectives analytically, or even proving their existence, is an open research problem \cite{adem2012symmetry, lukashchuk2015conservation, adem2015conservation, el2016variational, el2018modulational, el2017new, el2017stability}; thus, methods for deriving the differential equations of a system leave this question unsolved.

The dynamic reference value tightly followed in the observed system, might result from an environmental input, such as circadian rhythms. IDRAS, by decoupling the regulated combination from the dynamics of the reference value, yields a robust combination that is invariant in different environments. This notion resembles the objective of Invariant-Risk-Minimization defined in \cite{arjovsky2019invariant} for a classification problem, where some features in the data are environment-dependent and some are invariant. 


%
%

\newpage

%
%

\vskip 40pt
{\Huge{\textbf{Supplementary material}}}

\renewcommand{\thesection}{SI\arabic{section}}\setcounter{section}{0}   

\renewcommand{\thefigure}{SI~\arabic{figure}}
\setcounter{figure}{0}
\renewcommand{\thetable}{SI~\arabic{table}}
\setcounter{table}{0}
\renewcommand{\theequation}{S.\arabic{equation}}
\section{Validation}
\subsection{Architecture}

All the parameterized functions used in Section 3 are feed-forward artificial neural networks with an input layer of 32 neurons, a hidden layer of 16 neurons and an output layer of appropriate dimension. The activation function of all neurons is Leaky-ReLU except for the output neuron whose activation function is the Sigmoid function. The hyper-parameters are $T=10$ and $n_y=2$. We train the networks as described in Algorithm 1 for a pre-defined fixed number of epochs ($150$) using the Stochastic-Gradient-descent optimizer with a momentum value of $0.9$ and a learning rate of $0.01$. See \cite{zhang2020dive} for a detailed explanation of feed-forward neural networks, activation functions and optimizers.

\subsection{A kinetic model of
interactions}
Following are the exact details of the parameters used in the kinetic model of regulatory interactions validation example, Section 3.1.

The timescale of the feedback loop $\tau_{(P+S)}=\nicefrac{1}{f}=0.0005(F=2000)$ is much shorter than the timescale of environmental influenced oscillations in $K$, $\tau_K=0.2$. We simulated a dataset of $100$ observed systems sampled at rate $f_s=1$ with $K_s=K_p=150$, $\gamma_S=\gamma_P=70$, $\gamma_M=80$, $f=2000$, $K_0=300$ and $dW_P, dW_S \sim \mathcal{N}(0,(0.5dt)^2)$. For each system $\phi_K \sim \mathcal{U}[0,\pi]$ and $M(0),P(0),S(0) \sim \mathcal{U}[0.02,0.1]$.

\subsection{Bacterial life cycle}
Following are the exact details of the parameters used in the Bacterial life cycle validation example, Section 3.2.

The simulated dataset contains $100$ lineages each consisting of $100$ generations.In Eq. 17, $\mu_u=1\, \mathrm{[\mu m]}$, $\sigma_u=0.1\, \mathrm{[\mu m]}$ and $\tau_u=200\, \mathrm{[min]}$; Initial conditions are $u(0) \sim N(\mu_u, \sigma_u^2)$ for each lineage. In Eq. 18, the growth rate is sampled from a Gamma-distribution, $\alpha_n \sim \Gamma(25,9.4\cdot 10^{-4})$ and the division fraction is distributed $f_n \sim {\mathcal N}(0.5,0.05^2)$. The initial conditions for the first cell in each lineage are $x_{0,b} \sim {\mathcal N}(0.5, 0.05^2)$, $t_{0,b} = 0$.

\bibliographystyle{IEEEtran}
\bibliography{ifacconf}

\end{document}